\documentclass[preprint,pre,showkeys,showpacs]{revtex4}
\usepackage{graphicx}
\usepackage{graphics}
\usepackage{latexsym}
\usepackage{subfigure}
\usepackage{multirow}

\begin{document}

\title{Non-universal dependence of spatiotemporal regularity on
  randomness in coupling connections}

\author{Zahera Jabeen} \affiliation{Institute
  of Mathematical Sciences, Chennai, India.}  \author{Sudeshna Sinha}
\affiliation{Institute of Mathematical Sciences, Chennai, India.}
 \date{\today} \pacs{}
\begin{abstract}
  We investigate the spatiotemporal dynamics of a network of coupled
  nonlinear oscillators, modeled by sine circle maps, with varying
  degrees of randomness in coupling connections. We show that the
  change in the basin of attraction of the spatiotemporal fixed point
  due to varying fraction of random links $p$, is crucially related to
  the nature of the local dynamics. Even the {\em qualitative}
  dependence of spatiotemporal regularity on $p$ changes drastically
  as the angular frequency of the oscillators change, ranging from
  monotonic increase or monotonic decrease, to non-monotonic
  variation. Thus it is evident here that the influence of random
  coupling connections on spatiotemporal order is highly
  non-universal, and depends very strongly on the nodal dynamics.
\end{abstract}
\maketitle

\section{Introduction}

The dynamics of spatially extended systems has been a focus of intense
research activity in the past two decades. In recent years it has
become evident that modeling large interactive systems by finite
dimensional lattices on one hand and fully random networks on the
other, is inadequate, as various networks, ranging from collaborations
of scientists to metabolic networks, do not fit in either paradigm
\cite{RMP_bar}. Some alternate scenarios have been suggested, and one
of the most popular ones is the small-world network \cite{Watts}.
Here one starts with a structure on a lattice, for instance regular
nearest neighbor connections. Then each link from a site to its
nearest neighbor is rewired randomly with probability $p$, {\it i.e.}
the site is connected to another randomly chosen lattice site. This
model is proposed to mimic real life situations in which non-local
connections exist along with predominantly local connections.

There is much evidence that random non-local connections, even in a
small fraction, significantly affects geometrical properties, like
characteristic path length \cite{Watts}. However its {\em
  implications for dynamical characteristics is still unclear and even
  conflicting}.  While the dynamics of coupled oscillators and coupled
maps on regular lattices (known as `coupled map lattices' or CMLs) has
been extensively investigated \cite{cml}, there have been far fewer
studies on the spatiotemporal dynamics of nonlinear elements on
networks of different topologies \cite{dyn_net}. Most studies so far
have indicated that the regularity of systems increase monotonically
with $p$ \cite{sinha}.

In this paper we will provide evidence of a system where the
dependence of spatiotemporal regularity on the degree of randomness in
coupling connections is highly {\em non-universal}. We will show how
this dependence ranges from {\em monotonically increasing} to {\em
  monotonically decreasing}, via {\em non-monotonic variation}, as the
local dynamics changes. 
Thus we will demonstrate that the interplay between local dynamics and
connectivity acts in non-trivial and non-intuitive ways, and so even
the qualitative effect of random links on spatiotemporal regularity
can be completely reversed by changing the nodal dynamics.

\section{The Model}  

Here we consider nonlinear oscillators coupled to nearest neighbors on
a regular ring, with some fraction $p$ of the regular links rewired
randomly.  The individual sites (nodes) are modeled by sine circle
maps, which have widespread relevance for oscillatory phenomena
\cite{sinecircle}, and are given as:
$$f(x)=x+\Omega-\frac{K}{2\pi} \sin (2\pi x)$$
wherein $K$ measures the strength of the nonlinear term, and $\Omega$
represents the natural frequency of the map in the absence of
nonlinearity (i.e.  when $K=0$).  We restrict our studies to the
parameter region: $0\leq\Omega\leq \frac{1}{2\pi}$ and $K=1$. In this
region, the single sine circle map settles down to the spatiotemporal
fixed point, $x^{\star}=\frac{1}{2\pi} \sin^{-1}(\frac{2\pi\Omega}{K})$.

Under diffusive coupling, such a coupled sine-circle map lattice, is
given as:
\begin{equation}
x_{n+1} (i) = (1-\epsilon) f(x_n(i)) + \frac{\epsilon}{2} [f(x_n(i-1)) + f(x_n(i+1))] \pmod{1}
\end{equation}
where $i=1,\ldots L$ denotes the site index, $n$ denotes the time
index, and $\epsilon$ represents the coupling strength between the
sites ($0\leq\epsilon\leq 1$). Periodic boundary conditions have been
used, namely one has a ring of oscillators.

Earlier studies of this coupled map lattice have been carried out for
various types of initial conditions \cite{sinecml1}. In particular,
the evolution of this coupled map lattice with random initial
conditions shows interesting spatiotemporal dynamics including
spatiotemporal fixed points, spatial and spatiotemporal intermittency,
and spatiotemporal chaos \cite{sinecml2}.

Now we introduce randomness in this regular lattice by rewiring the
nearest neighbor links with a probability $p$ to randomly chosen
sites on the lattice. Here we consider the behavior of the system
under {\em static rewiring}, namely, the randomness in spatial
coupling is quenched or ``frozen'' in time. Ensembles of such randomly
rewired systems are studied.

\section{Results}

\subsection{Basin of attraction for the spatiotemporal fixed point}

\begin{figure}[!t]
\begin{center}
\begin{tabular}{ll}
\includegraphics[height=1.8in,width=2.7in]{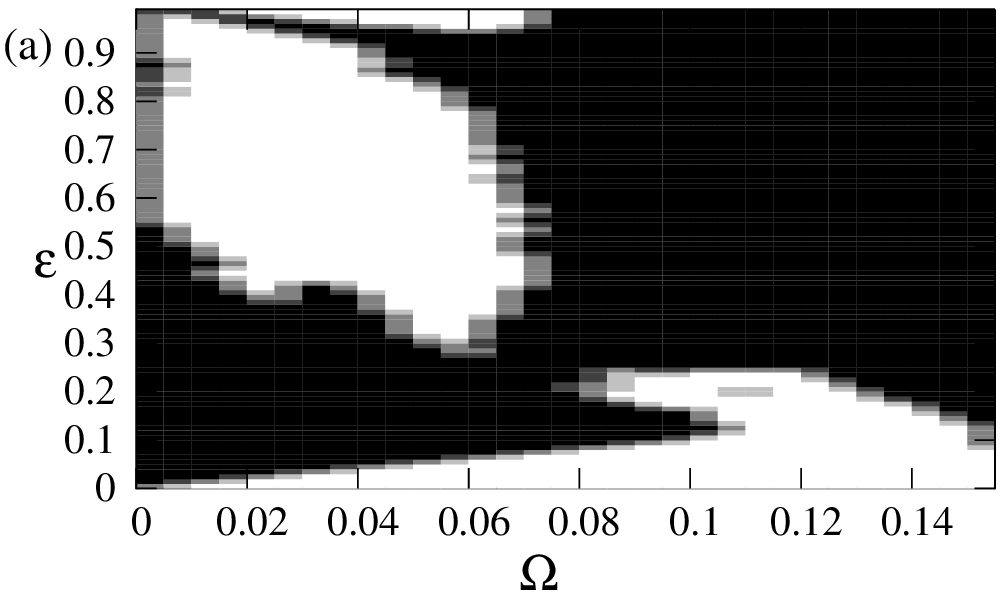}& \includegraphics[height=1.8in,width=3.0in]{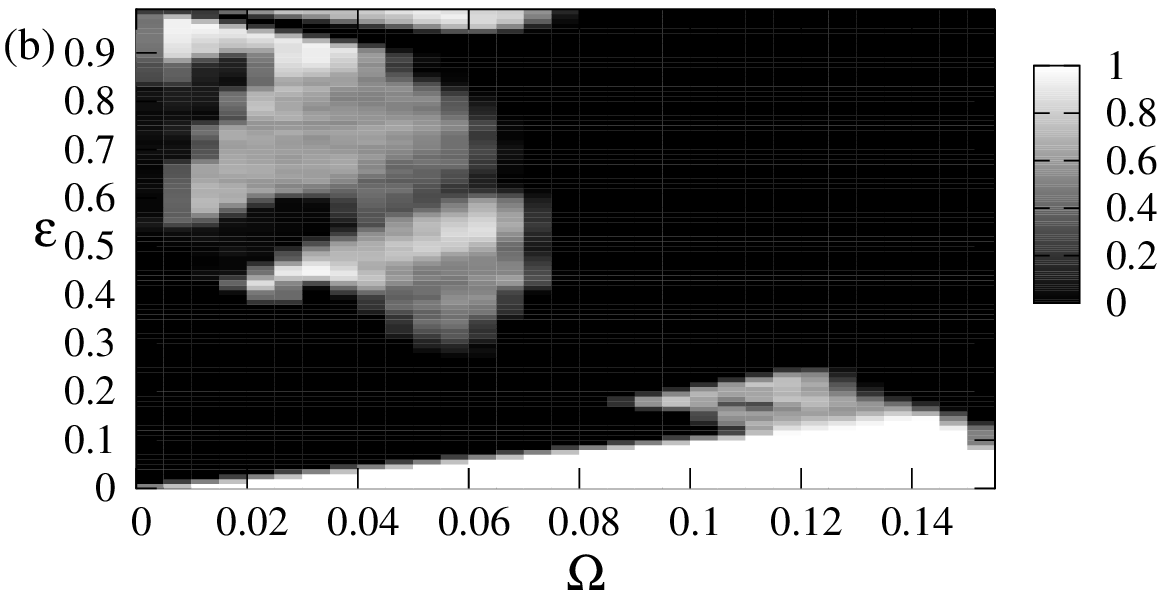}\\
\includegraphics[height=1.8in,width=2.7in]{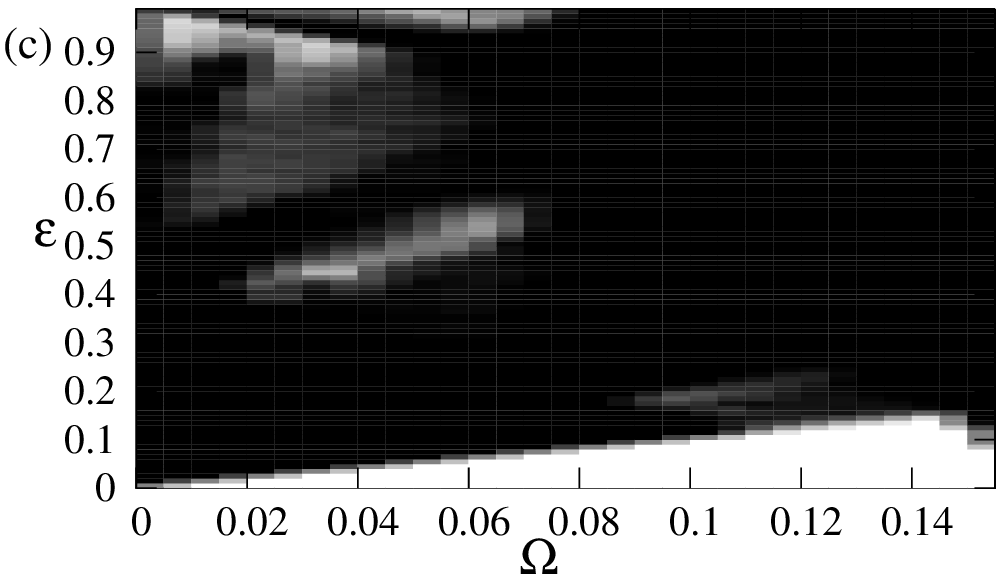}& \includegraphics[height=1.8in,width=3.0in]{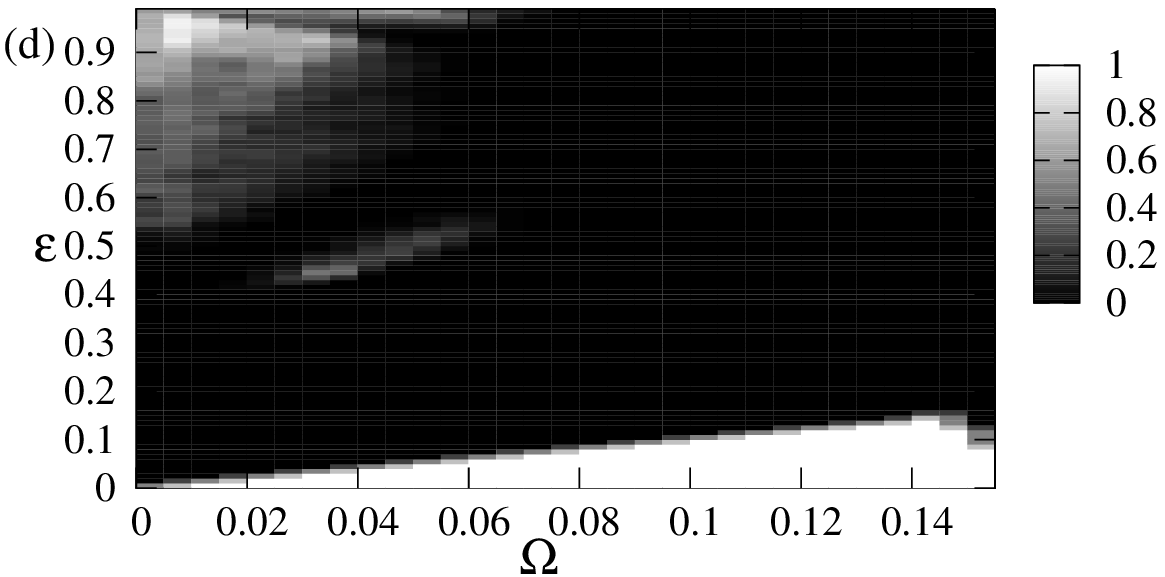}\\
\includegraphics[height=1.8in,width=2.7in]{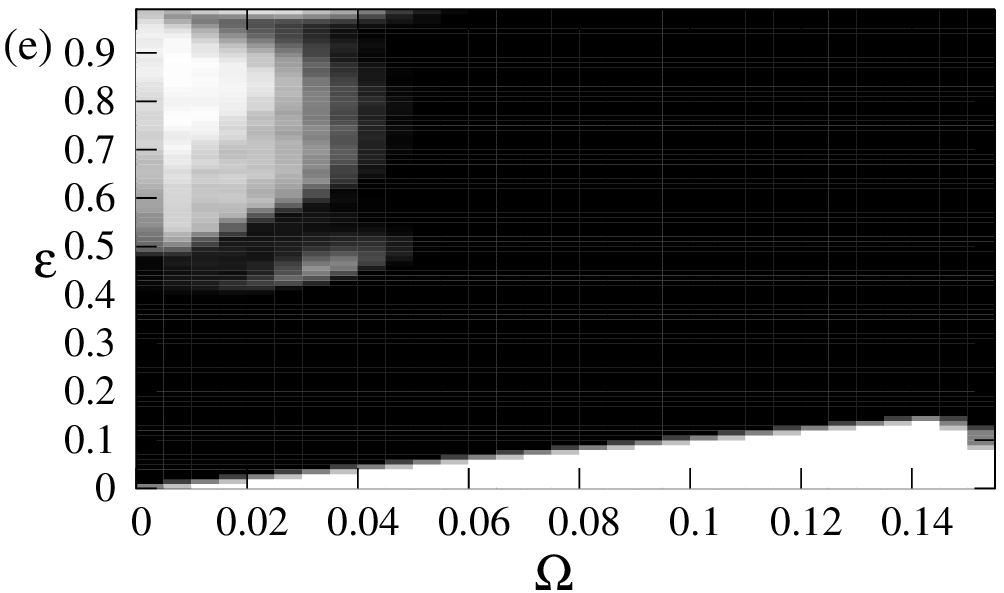}& \includegraphics[height=1.8in,width=3.0in]{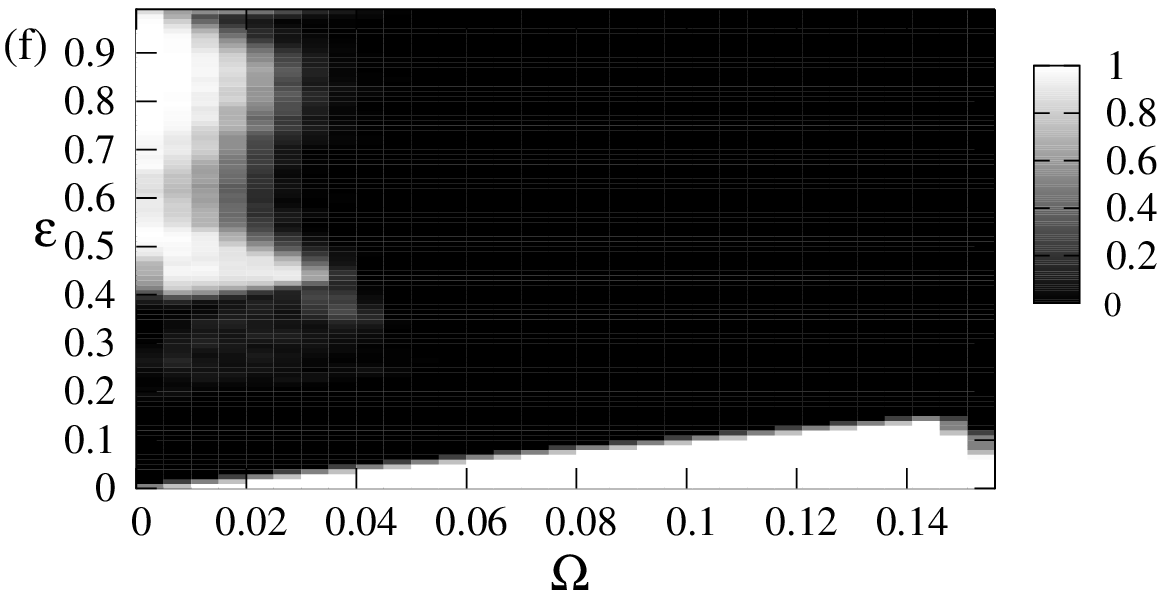}\\\end{tabular}
\end{center}
\caption{Basin of attraction of the spatiotemporal fixed point,
  $x^{\star}$, obtained for different random rewiring probabilities: (a) $p=0.0$, (b)
  $p=0.04$ (c) $p=0.1$, (d) $p=0.3$, (e) $p=0.5$, and (f) $ p=0.8$.
  These plots have been obtained for $50$ rewiring configurations for
  a lattice of size $L=200$ after discarding $5000$ transients.
  \label{basin}}
\end{figure}

We study the spatiotemporal dynamics of this system, starting from
random initial conditions under varying rewiring probabilities $p$ ($0
\le p \le 1$), for different realizations of static rewiring.

Specifically we obtain the basin of attraction for the spatiotemporal
fixed point, $B$, by calculating the fraction of rewiring
configurations which lead to a spatiotemporally steady state for
different ($\Omega,\epsilon$) values. Figure \ref{basin} shows a gray
scale plot of $B$ in a large region of parameter space, for different
values of $p$. The white areas indicate the parameter regions, where
all initial coupling configurations lead to a spatiotemporal fixed
point, namely all the sites in the system relax to the fixed point
$x^{\star}$ such that $x(i)=x^{\star}$ for all $i=1\ldots N$ and for
all time $n$. The black regions indicate parameter regions where none
of the coupling configurations yield a spatiotemporal fixed pont. The
gray areas indicate parameter regimes where $0 < B < 1$, i.e. the
spatiotemporal fixed point co-exists with other dynamical behaviour,
and the spatiotemporal fixed point is not an attractor of the dynamics
for all coupling configurations.

\begin{figure}[!t]
\begin{center}
\begin{tabular}{ll}
\includegraphics[height=1.8in,width=2.7in]{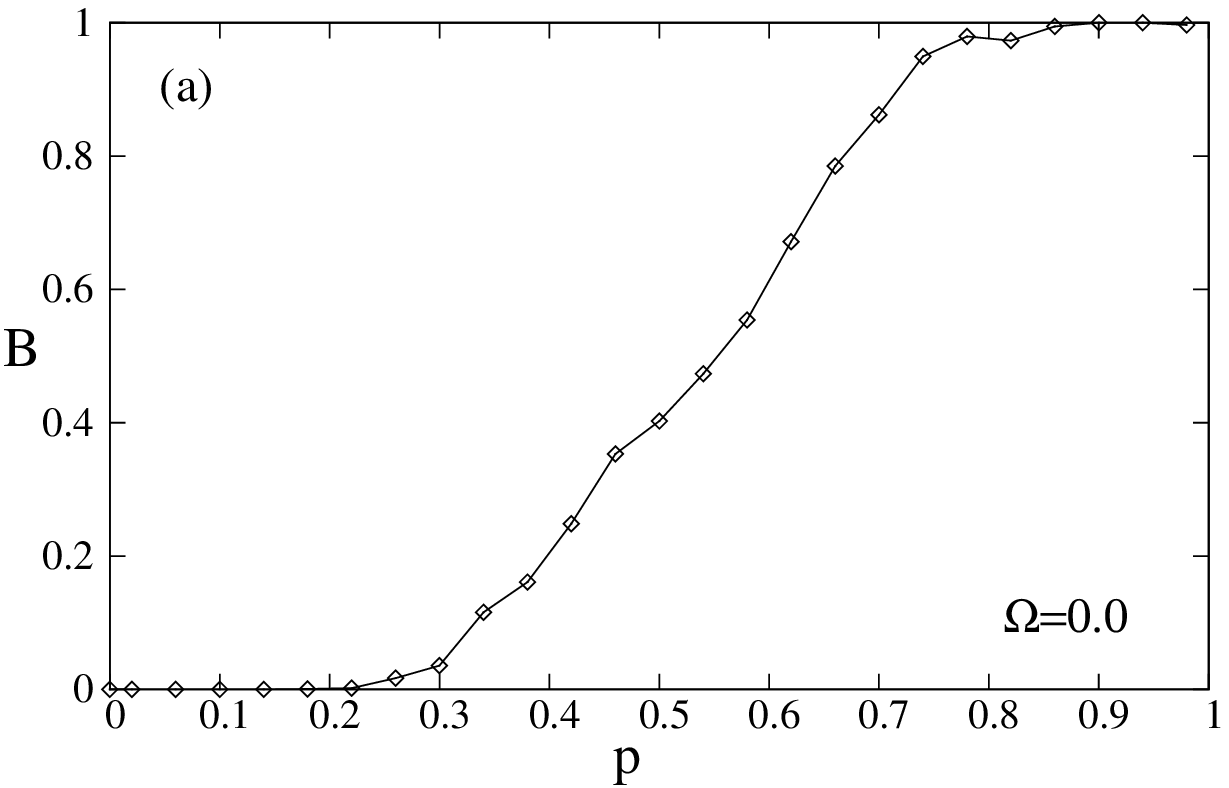}& \includegraphics[height=1.8in,width=2.7in]{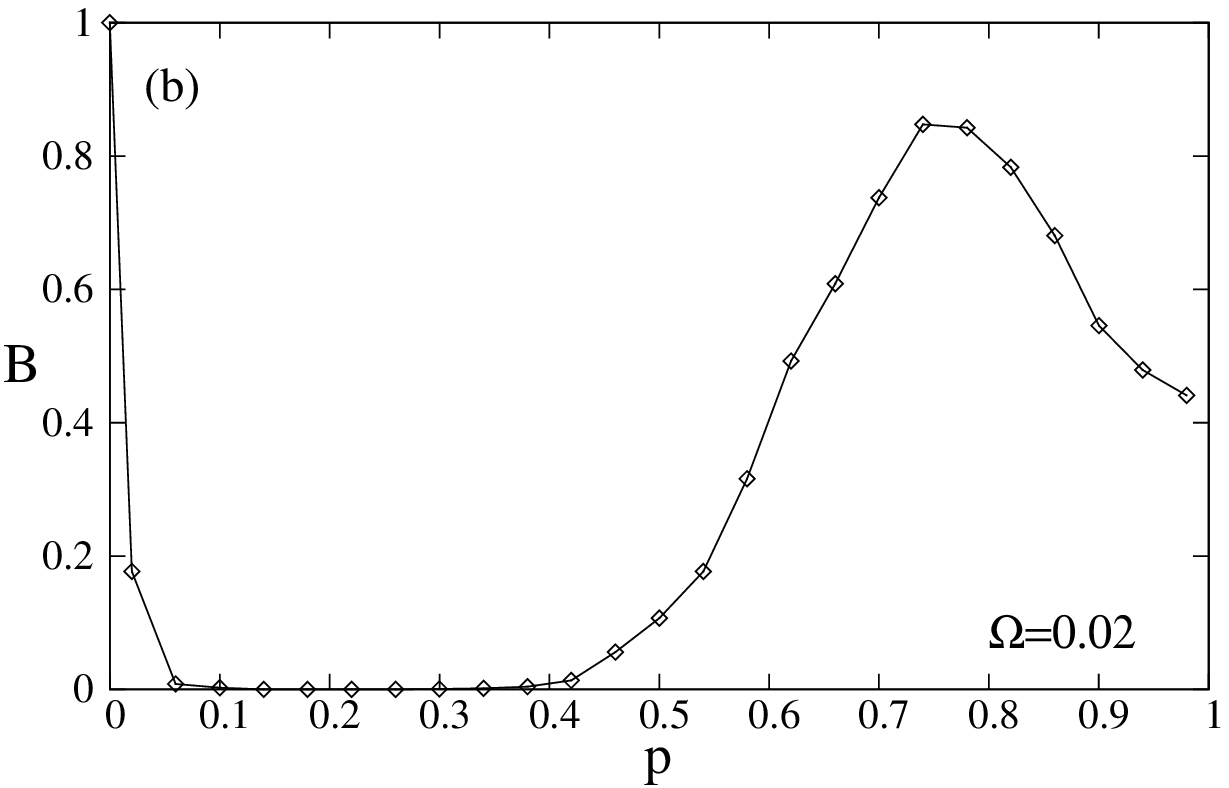}\\
\includegraphics[height=1.8in,width=2.7in]{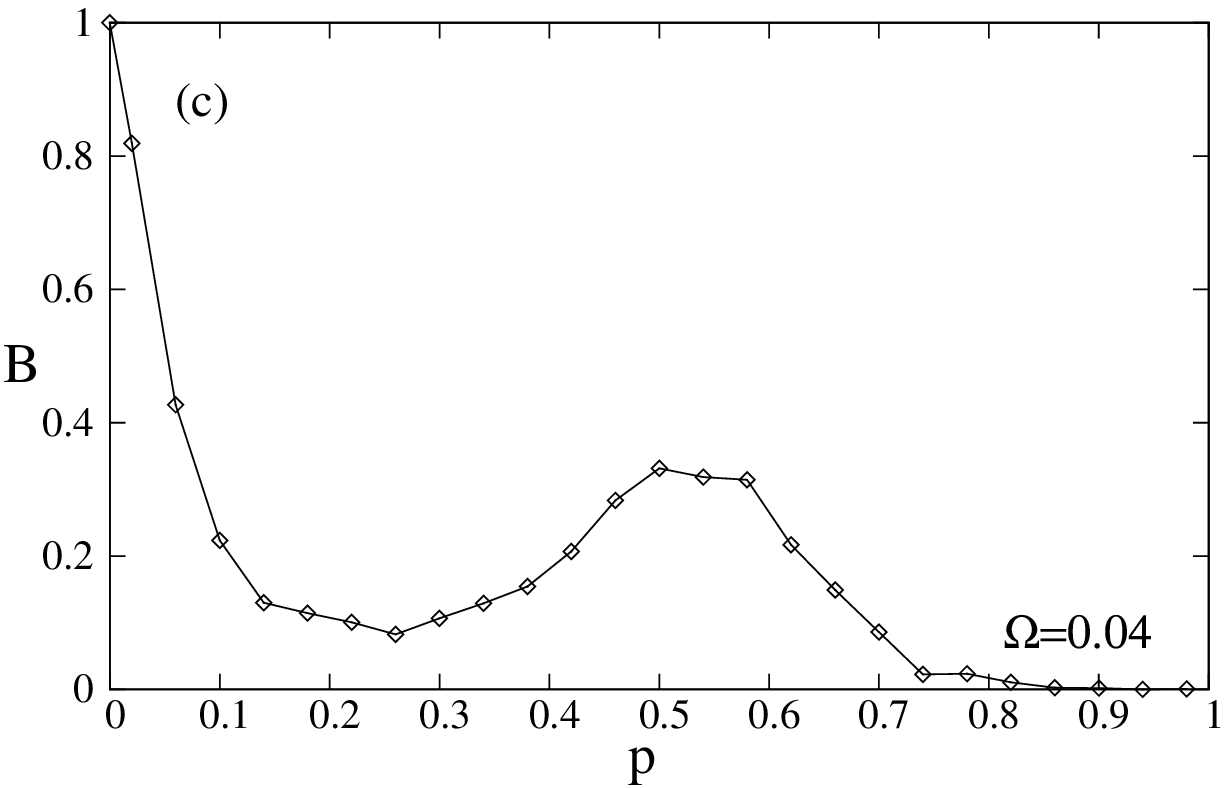}& \includegraphics[height=1.8in,width=2.7in]{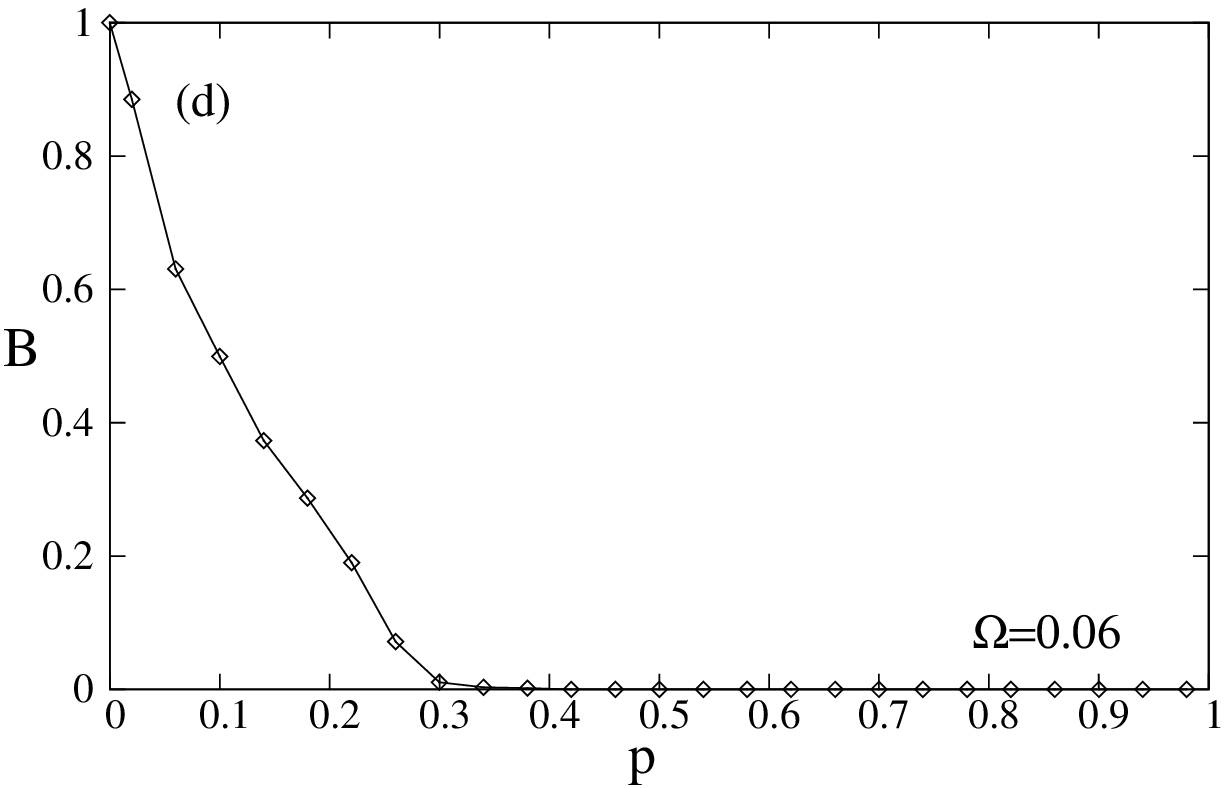}\\
\end{tabular}
\end{center}
\caption{ Variation of basin of attraction $B$ with the
  rewiring fraction, $p$ plotted for the circle map frequencies
  $\Omega=0, 0.02, 0.04, 0.06$ and coupling strength $\epsilon=0.5$.
  \label{e5ovar}}

\end{figure}

Figure \ref{basin}(a) shows the basin of attraction when the rewiring
fraction is equal to zero, or in other words, the ring has only
regular nearest neighbor connections. As the rewiring fraction $p$ is
varied, the spatiotemporal fixed point regions also show a change.
Figure \ref{basin}(b)-(f) display the basin of attraction of the
spatiotemporal fixed point, for $p=0.04, 0.1, 0.3, 0.5$ and $p=0.8$.
We see that the dependence of this basin of attraction on the degree
of random rewiring is qualitatively very different for different
values of $\Omega$ and $\epsilon$.  Interestingly this variation
ranges from monotonic increase to monotonic decrease, as well as
non-monotonic behavior, along different ``cuts'' in
($\Omega,\epsilon$) space.

Further, the dependence of spatiotemporal order on rewiring
probability, averaged over a large parameter range, varies
non-monotonically with $p$. This is clear from the fact that the
extent of the spatiotemporal fixed point basin at intermediate $p$ ($p
\sim 0.1-0.3$) in Figs.\ref{basin}(c-d) is much smaller than that for
low and high $p$. So the gray-scale basin plots in Figs.\ref{basin}
(c-d) appear far less ``white'' in general, across large parameter
regimes, as compared to Figs.\ref{basin} (a) and (f).

Figure \ref{e5ovar} shows the variation of the basin of attraction
$B$ with rewiring fraction $p$ at $\epsilon=0.5$ and for
$\Omega=0.0, 0.02, 0.04,$ and for $0.06$. These plots have been
obtained for $50$ rewiring configurations for a lattice of size
$L=200$ after discarding $10000$ transients.  When the natural
frequency of the circle map is equal to zero, ($\Omega=0$), the ring
does not yield a spatiotemporal fixed point when coupling connections
are completely regular.  However, the regularity of the system
increases as the rewiring fraction $p$ is increased. This can be seen
in Figure \ref{e5ovar}(a), in which a global spatiotemporal fixed
point attractor is obtained for values of the rewiring fraction $p >
0.6$.

\begin{figure}[!t]
\begin{center}
\begin{tabular}{ll}
\includegraphics[height=1.8in,width=2.7in]{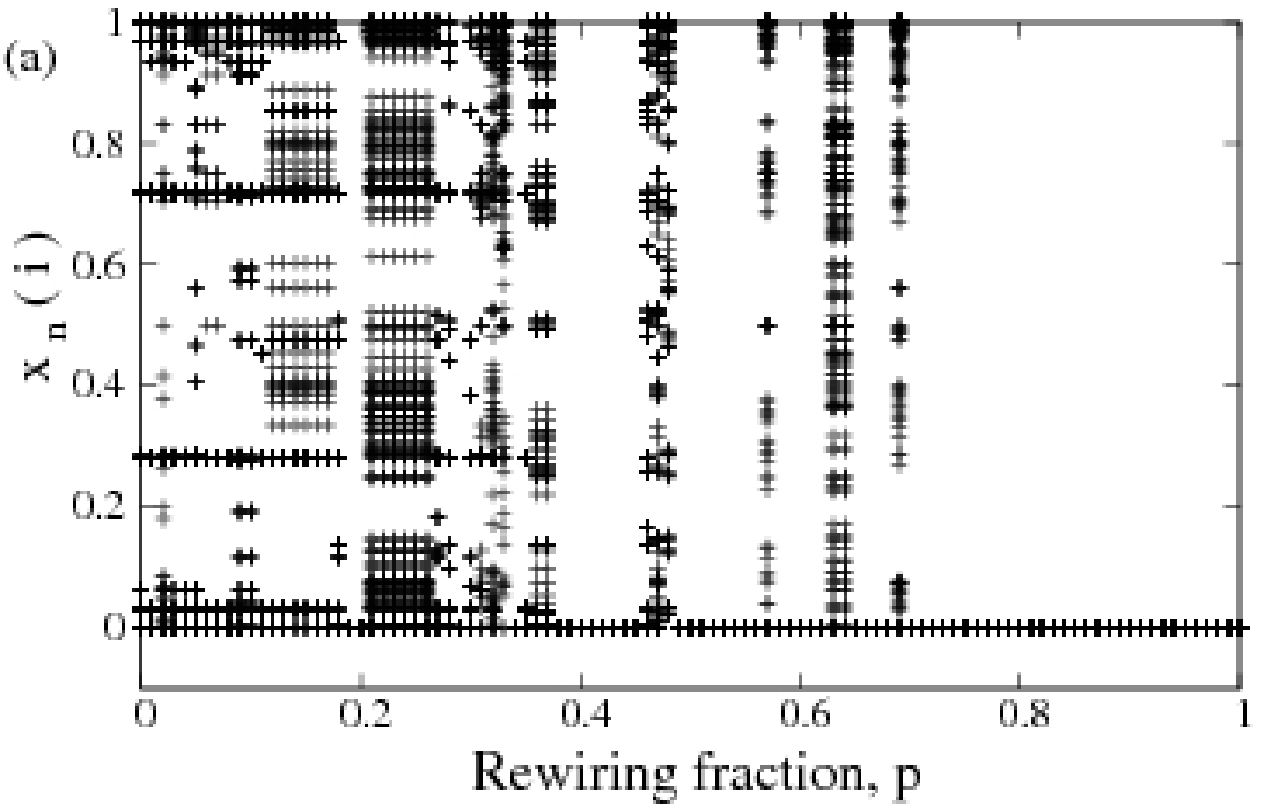}& \includegraphics[height=1.8in,width=2.7in]{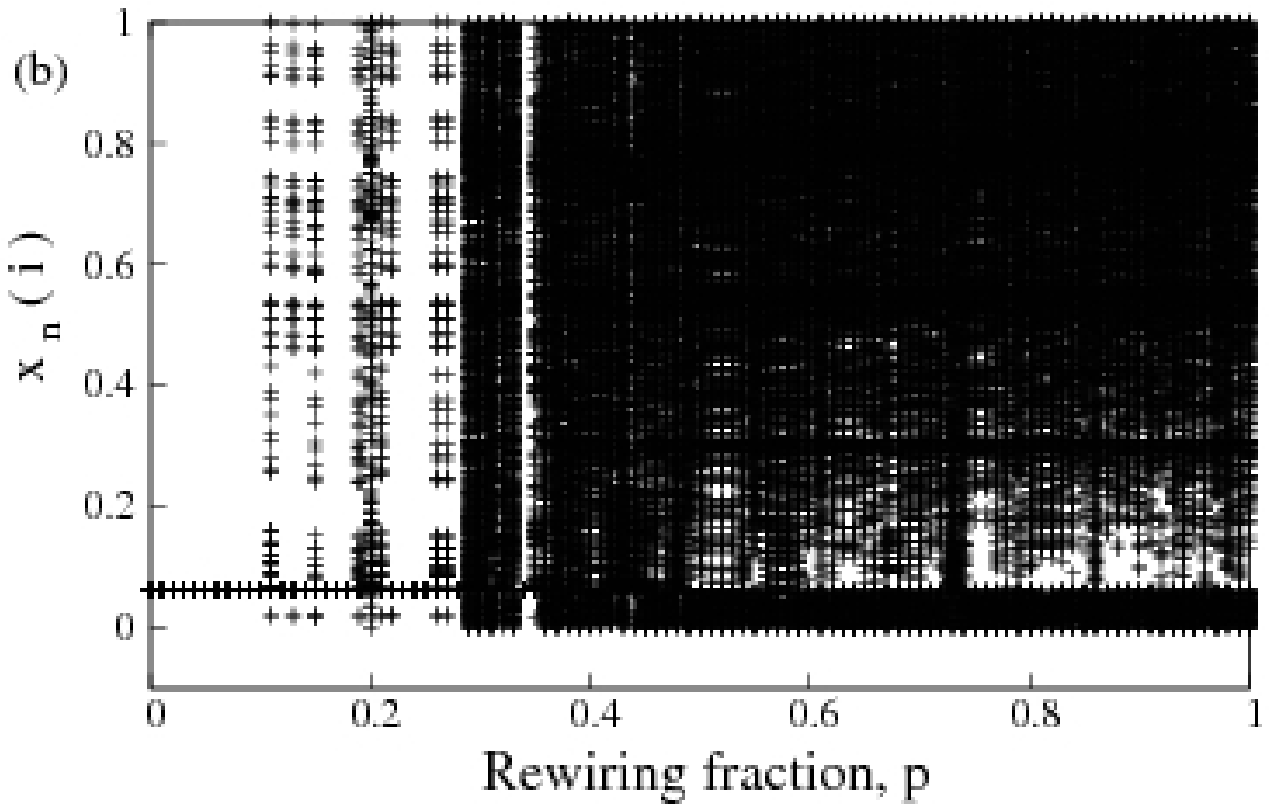}\\
\end{tabular}
\end{center}
\caption{ Bifurcation diagram in which the state variables $x_n(i)$
  ($i=1,\ldots,100$) have been plotted as a function of the fraction
  of random links, $p$, for a representative coupling configuration,
  for the parameter values (a) $\Omega=0.0$ and (b) $\Omega=0.06$ at
  $\epsilon=0.5$. Here, $n=1,\ldots,5$ iterations have been plotted
  after discarding $5000$ transients. Note that for regions where $0 <
  B < 1$, there are rewiring configurations that lead to the
  spatiotemporal fixed point, co-existing with rewiring configurations
  that yield spatiotemporal chaos. \label{bifur}}
\end{figure}

The bifurcation diagram in Figure \ref{bifur}(a), showing the
spatiotemporal dynamics of the system with respect to the fraction of
random links $p$, further underscores this feature. Here the system
has a complex spatial pattern for lower values of random rewiring
probability. However, it settles down to the spatiotemporal fixed
point ($x^{\star}=0$, in this case) as the fraction of random links
$p$ approaches $1$.

In contrast, the variation of the basin of attraction, $B$, in
the case where the frequency $\Omega$ is equal to $0.02$ is shown in
Figure \ref{e5ovar}(b). Here, we see that the system yields a
spatiotemporal fixed point with probability $1$ for zero rewiring
fraction $p$, but shows a non-monotonic variation as the rewiring
fraction $p$ is changed. We see that though the basin of attraction
decreases to zero in the interval $p\sim 0.1-0.4$, it gradually
increases for rewiring fractions, $p>0.4$, until it again registers a
decrease in the large $p$ limit \cite{foot}. Hence, the basin of
attraction $B$ shows a {\em non-monotonic} variation with change in
$p$.  A similar non-monotonic variation is seen in Figure
\ref{e5ovar}(c), where $\Omega=0.04$.

\begin{figure}[!t]
\begin{center}
\begin{tabular}{ll}
\includegraphics[height=1.8in,width=2.7in]{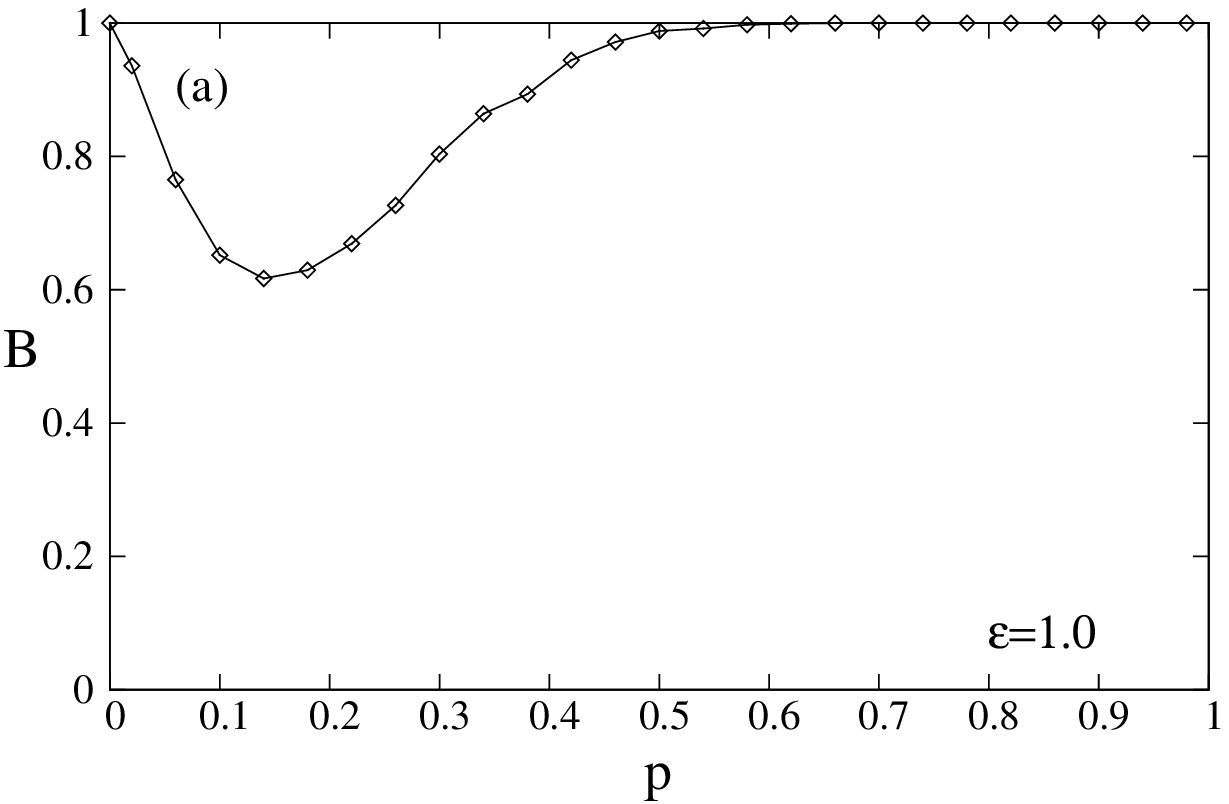}& \includegraphics[height=1.8in,width=2.7in]{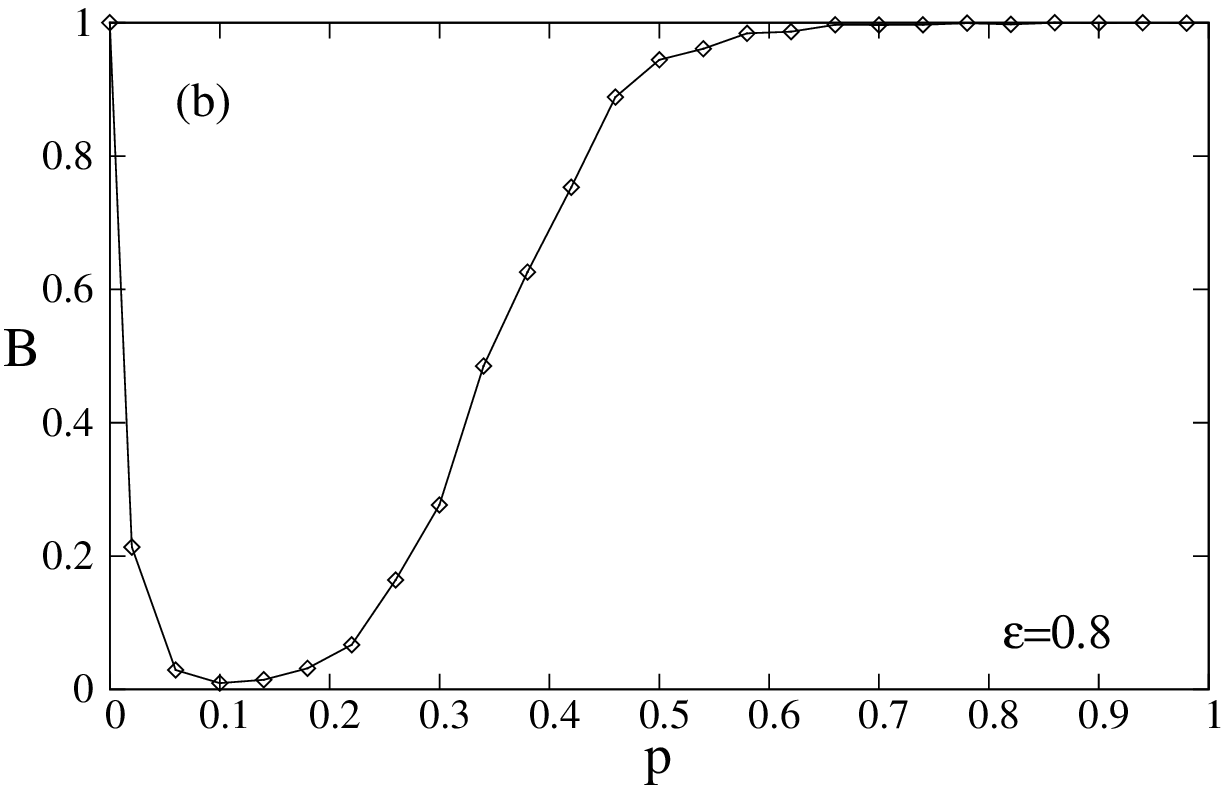}\\
\includegraphics[height=1.8in,width=2.7in]{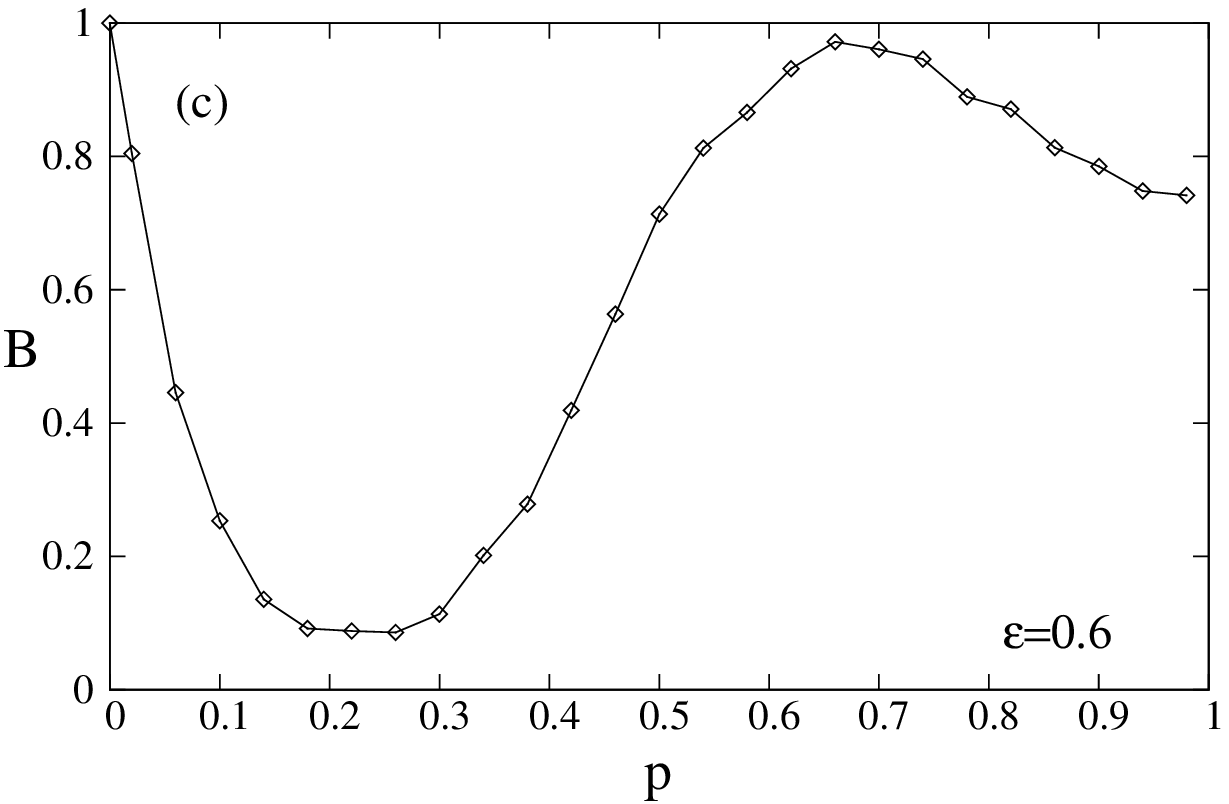}& \includegraphics[height=1.8in,width=2.7in]{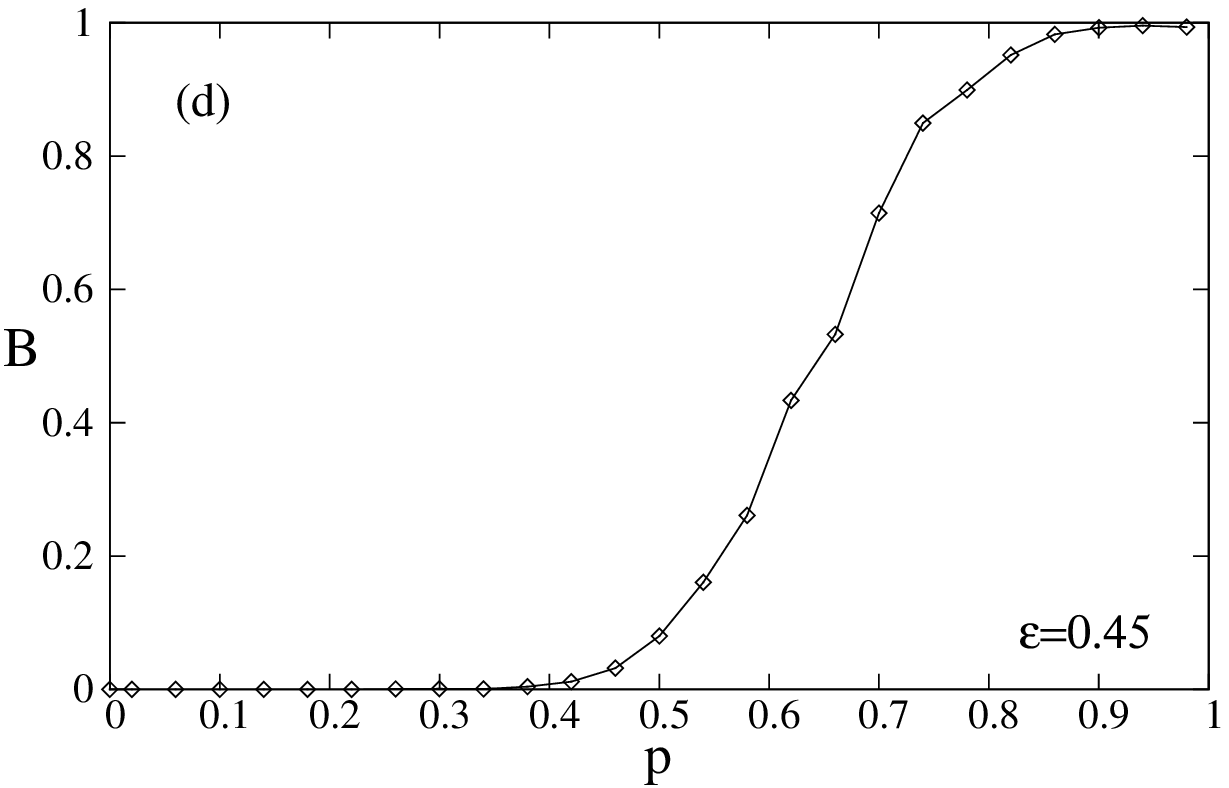}\\
\end{tabular}
\end{center}
\caption{ Variation of basin of attraction, $B$ with the
  rewiring fraction, $p$ for the circle map frequency, $\Omega=0.01$
  and coupling strengths, $\epsilon=1.0, 0.8, 0.6,$ and $0.45$.
  \label{o01evar}}

\end{figure}

In the case of $\Omega=0.06$, as displayed in Figure \ref{e5ovar}(d),
the basin of attraction {\em decreases} to zero as the rewiring
fraction is increased. In this case, the system settles to the
spatiotemporal fixed point $x^{\star}$ for smaller rewiring fractions,
but exhibits a complex spatial pattern when the degree of rewiring in
the system is increased. This is further illustrated in the
bifurcation diagram of the system shown in Figure \ref{bifur}(b). This
is exactly the opposite trend to that observed in the case of
$\Omega=0$. So, as the local frequency of the nonlinear oscillator
changes, the effect of random rewiring on spatiotemporal properties is
completely reversed.

Hence, we see that for the same coupling strength $\epsilon$, and for
the same set of rewired configurations, the basin of attraction of the
spatiotemporal fixed point shows a very strong dependence on the local
dynamics, namely on the frequency $\Omega$ of the nonlinear
oscillators. So it is evident that the spatiotemporal regularity
depends crucially, not just quantitatively, but also qualitatively, on
the nodal dynamics.

Similarly, when the nodal dynamics is fixed and the coupling strength
$\epsilon$ is varied, we see that the basin of attraction shows a
non-monotonic variation with change in rewiring fraction $p$. This is
illustrated in Figure \ref{o01evar} where the basin of attraction has
been plotted for $\Omega=0.01$ and for various representative values
of the coupling strength $\epsilon$.

Hence, spatiotemporal regularity of the dynamics on a network depends
quite crucially on the interplay between the nodal dynamics and the
network topology. That is, coupling configurations with the same
degree of randomness may enhance or inhibit spatiotemporal order
depending on the properties of the local oscillators.

\section{Conclusions}

In summary, we have investigated the spatiotemporal dynamics of a
network of coupled nonlinear oscillators, modeled by sine circle maps,
with varying degrees of randomness in coupling connections. We showed
that the variation of the basin of attraction of the spatiotemporal
fixed point, with increasing fraction of random links $p$, crucially
depends on the nature of the local dynamics. Even the qualitative
relationship between spatiotemporal regularity and $p$ changes
drastically as the angular frequency of the oscillators change,
ranging from monotonic increase or decrease, to non-monotonic
variation. Thus it is evident that the influence of random coupling
connections on spatiotemporal order is highly non-universal here and
depends strongly on the angular frequency of the nodal oscillators. 
This implies that the delicate interplay between local dynamics and
connectivity is crucial in determining the emergence of spatiotemporal
order in complex networks of dynamical elements.\\

{\bf Acknowledgement:} We would like to thank Prof. N. Gupte for
useful discussions.



\begin{thebibliography}{99}
\bibitem{foot} Note that the average time taken by the system to relax
  to the spatiotemporal fixed point in the large $p$ regime ($p\sim
  0.8-1.0$), grows exponentially with $p$, and the configurations
  eventually relax to the spatiotemporal fixed point with probability
  $1$, but only at asymptotically long times. This is in contrast to
  the region of rewiring fractions $p\sim 0.1-0.4$, where most initial
  states do not appear to relax to the spatiotemporal fixed point
  state, even after very long transience.
\end{thebibliography}
\end{document}